\documentclass[reprint,superscriptaddress,prb]{revtex4-1}

\usepackage{graphicx}
\usepackage{amsmath}

\begin{document}

\title{Species Fractionation in Atomic Chains from 
Mechanically Stretched Alloys}

\author{Pedro Alves da Silva Autreto}
\affiliation{Institute of Physics Gleb Wataghin,
             University of Campinas, UNICAMP
             Cidade Universitaria, 13083-970, Campinas, SP, Brazil}
\affiliation{Department of Earth Sciences, University of Cambridge, 
             Cambridge CB2 3EQ, United Kingdom}
\author{Douglas S. Galvao}
\affiliation{Institute of Physics Gleb Wataghin,
             University of Campinas, UNICAMP
             Cidade Universitaria, 13083-970, Campinas, SP, Brazil}
\author{Emilio Artacho}
\affiliation{Department of Earth Sciences, University of Cambridge, 
             Cambridge CB2 3EQ, United Kingdom}
\affiliation{Theory of Condensed Matter, Cavendish Laboratory, 
University of Cambridge, Cambridge CB3 0HE, United Kingdom}
\affiliation{Nanogune and DIPC, 
Tolosa Hiribidea 76, 20018 San Sebasti\'an, Spain}
\affiliation{Basque Foundation for Science, Ikerbasque, 
48011 Bilbao, Spain}

\date{\today}

\begin{abstract}
 Bettini \textit{et al.} [Nature Nanotech {\bf 1}, 
182 (2006)] reported the first experimental realization of linear 
atomic chains (LACs) composed of different atoms (Au and Ag). 
  Different contents of Au and Ag were observed in the chains from
what found in the bulk alloys, which rises the question of what is the 
wire composition if in equilibrium with a bulk alloy. 
  In this work we address the thermodynamic driving force for 
species fractionation in LACs under tension, and we present
density-functional theory results for Ag-Au chain alloys.
  A pronounced stabilization of wires with an alternating Ag-Au 
sequence is observed, which could be behind the experimentally 
observed Au enrichment in LACs from alloys of high Ag content.
\end{abstract}

\maketitle

\section{Introduction}
  Metal nanowires have attracted great interest with the increasing 
miniaturization of electronic and mechanical devices \cite{Jia2007},
given their very interesting properties related to quantum 
transport \cite{Agrait2003}.
  From the experimental point to view, two techniques have been 
mostly used in nanowires research: mechanically controllable break 
junctions (MCBJ) \cite{Yanson1999, Mares2005, Rodrigues2000} 
and \textit{in situ} high-resolution transmission electron microscopy 
(HRTEM) \cite{Kondo1997,Kizuka1998, Kizuka1997, Koizumi2001, 
Ohnishi1998, Rodrigues2000, Rodrigues2001}.
  While MCBJ is used to obtain electronic properties of nanowires, 
HRTEM gives structural information, providing atomic resolution 
with real-time image acquisition. 
  A combination of both techniques is frequently used in order to 
establish a correlation between atomic arrangement and quantum conductance \cite{lagos2012correlation}. 
  Since 1997 these techniques have revealed the existence of some 
quite unexpected structures, such as, extremely thin metal nanotubes 
with a square cross-section \cite{Lagos2009}, helical multishell gold 
nanowires \cite{Kondo2000}, single-walled metal nanotubes 
\cite{Oshima2002, Oshima2003}, and the ultimate nanowires, 
linear atomic chains (LACs)  \cite{Ohnishi1998, coura2004structural}.
  The stability of these structures has been inferred using statistics over 
large number of observed structures (in HRTEM) and on 
conductance histograms for a large number of contact-breaking 
events (in MCJB).  
 These experimental analyses have shown the remarkable 
existence of stable (or metastable) nanowires for alkali 
\cite{Yanson1999} and noble metals.\cite{Mares2005}

\begin{figure}[b]		
\includegraphics[scale=0.4]{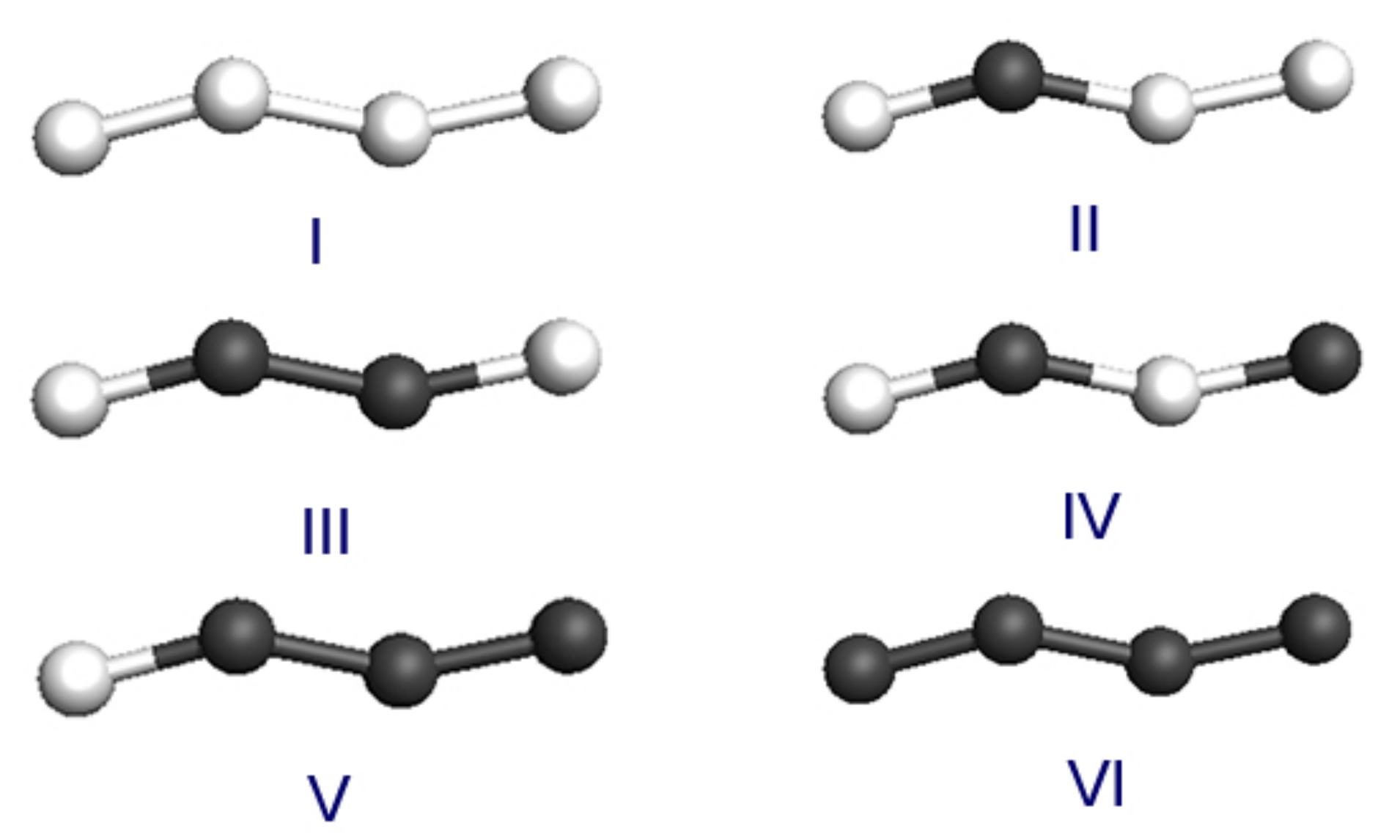}
\caption{\label{studiedgeometries} 
Structural motifs for the atomic-wire compositions considered in this work. 
Dark (light) balls indicate Ag (Au) atoms. See text for discussions.}
\end{figure}

  There have been many theoretical studies of these structures, 
mostly using classical atomistic models, as well as first principles 
ones \cite{nakamura1999density, sanchez1999stiff, Lagos2009,senger2004chiral}. 
  Of particular importance is the work of Tosatti {\it et al.} 
\cite{tosatti2001string}, that investigated the stability of these 
nanostructures considering the metal wires connected to atomic 
reservoirs at both ends, determining the local minima (different 
metastable structures) as a function of the tension of the wire.
  Using this approach the authors were able to predict the stability of 
magic sizes for Au nanowires. 

  Although metal nanowires have been the object of intense 
theoretical and experimental investigations, most of these studies 
were carried out for pure metals. 
  The first LAC experimental realization using metallic alloys
was reported only in 2006 \cite{Bettini2006}.
  They reported the observation of LACs of different lengths
composed of Ag and Au atoms from Ag-Au nanoalloys of 
different compositions.
  They observed that when the nanowires are rich in Ag 
\cite{Bettini2006}, in the final states of the stretching 
leading to LAC formation, the observed behavior was 
similar to the ones for Au rich nanowires \cite{Bettini2006}.
  From classical molecular dynamics \cite{Bettini2006} it was 
concluded that this unusual behavior is consequence of the 
Au enrichment at the region where the LACs were formed. 
  The concentration of these nanoalloys changes 
significantly in time from the nominal initial values. 
  They basically behave as if they had a much high Au content 
than in their nominal composition.  This behavior was addressed by Fao \cite{fa2008stability} 
by considering the enthalpy as defining the stability of these atomic chains under tension, in line with what pioneered by Tosatti \cite{tosatti2001string}. A grand canonical perspective, however, which is the adequate framework to study the thermodynamical trends in these systems, has not been done until this date and will be the focus of this work.

  The main objective of this paper is to determine the
relative concentrations of two species (Ag and Au in this case)
in LAC when in equilibrium with bulk alloys of given composition.
  For the purpose the relevant free energy is defined and 
calculated from first principles, analogously to what was 
done in ref. \cite{tosatti2001string} for single-species 
wires and tubes, which is now generalized to a 
grand-canonical setting. The observed capability of the wires to change composition with respect
to the bulk reservoir demands such grand canonical treatment, beyond what
was presented in ref. \cite{fa2008stability}.

In this way, with our approach we should be able to predict for specific 
bulk alloy concentrations what would be the corresponding
likely concentrations for LACs formed by stretching a nanowire. 
It is important to remember, however, that this method takes an equilibrium viewpoint, which assumes 
atom exchange between reservoirs and wire, and should only 
be taken as a determination of equilibrium trends. 
The kinetics of the formation process will very probably
originate deviations from this limiting behavior. It is however important to
know what the thermodynamical trends are, which is what we address here.

\section{METHODOLOGY}
  Density-functional theory (DFT) calculations were carried out
for free-standing infinite nanowires by preparing 
arrays of sufficiently separated LACs in periodic boundary 
conditions.
  Tetragonal supercells are used with the wires aligned 
along the $x$-axis, with four atoms of Au or Ag
and binary alloys of them. 
  The considered geometries are presented in 
Fig.~ \ref{studiedgeometries}. 
  The LAC geometries are relaxed for fixed supercell size,
keeping the $y$ and $z$ sizes such that the minimum 
distance among wires is 15.0 \AA, while the $x$ 
direction was varied in increments of 0.05 \AA, from 
6.00 \AA\ until the wire snapped.
  We have considered LAC lengths between 
7 \AA\ and 11 \AA.

\begin{figure}[t]
\centering
\includegraphics[scale=0.27]{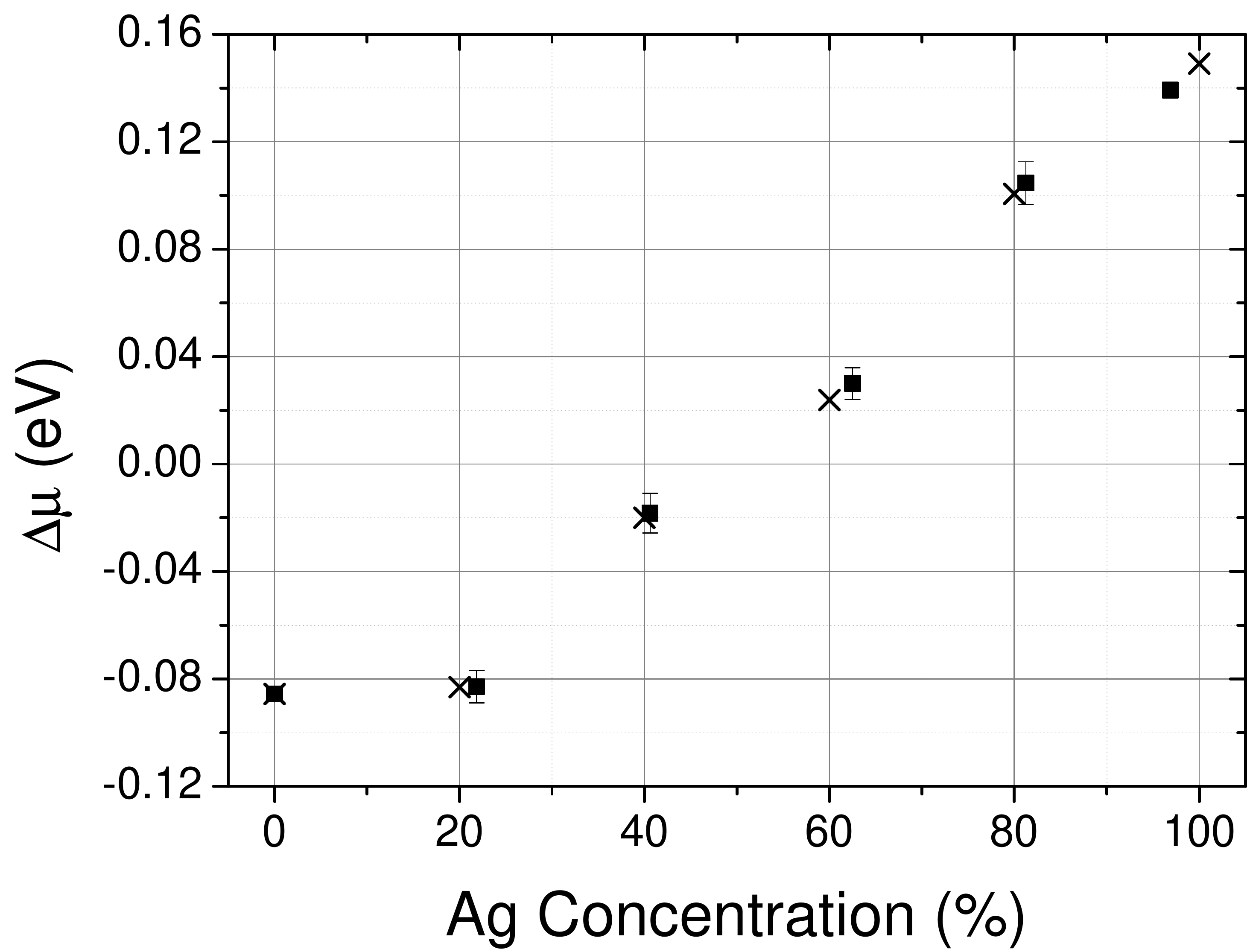}
\label{chemical}
\caption{\label{chemical} Chemical potential difference 
$\Delta \mu = (\mu_{\rm Au} - \mu_{\rm Ag})/2$ calculated 
for a bulk AgAu alloy. 
  Squares indicate $\Delta \mu$ values at bulk concentrations 
0, 7/32, 13/32, 20/32, 26/32 and 1; crosses show the interpolated 
estimates for concentrations 0, 0.2, 0.4, 0.6, 0.8 and 1.}
\end{figure}

  The DFT calculations were performed with the {\sc Siesta} 
method \cite{Ordejon1996, Soler2002}, using the 
Perdew-Burke-Ernzerhof (PBE) generalized gradient 
approximation (GGA) for exchange and 
correlation \cite{Perdew1996}. 
  Norm-conserving scalar-relativistic 
pseudopotentials \cite{Troullier1991} replaced nuclei 
and core electrons, considering the normal core-valence 
separation for Ag and Au. 
  Double-$\zeta$ polarized bases were used for valence 
electrons \cite{Anglada2002}. 
  Integrals in real space were performed on a mesh of 250 Ry 
cutoff \cite{Soler2002}. 
  The integration over the first Brillouin zone was carried out 
using the special $k$-points method \cite{Monkhorst1976} 
with $k$-meshes of $1 \times 1 \times 30$ and $10 \times 
10 \times 10$ for wire and bulk structures, respectively.

  In order to calculate the nanowire stability for an alloy we 
have generalized the string tension analysis to 
different atomic species. 
  The free energy $\Lambda$ we will seek to minimize is
\begin{equation}
\Lambda=E_{\rm LAC}-n_{\rm Au}\mu_{\rm Au}-n_{\rm Ag}\mu_{\rm Ag}
\end{equation}
where $n_{\rm Au}$ and $n_{\rm Ag}$ represent the number of 
Au and Ag atoms in the LAC (for the case of AgAu alloys), 
and $\mu_{\rm Au}$ and $\mu_{\rm Ag}$,  are the corresponding 
chemical potentials.
  It is the zero-temperature grand-canonical free energy 
(per unit cell) that gives the stability for a wire of unit-cell length 
$L$ (length of cell along $x$).
  The analysis can also be done as a function of $F$, the 
force pulling the wire, in which case the relevant free 
energy would be \cite{tosatti2001string}
\begin{equation}
\Theta = \Lambda - F L \;  . 
\end{equation}
  Here we will restrict ourselves to $L$-dependence and
thus $\Lambda$.
  We will consider the low temperature limit,
disregarding entropy effects.
  The generalization of the present scheme to finite temperature
is costly but conceptually straightforward, either by including the 
entropy of harmonic vibrations \cite{hobi2008temperature} 
(for a fully quantum albeit harmonic treatment), or thermodynamic 
integration for the free energy \cite{frenkel2002understanding},
using molecular dynamics for classical nuclei, or with path-integral
Monte-Carlo or molecular dynamics for quantum nuclei. 

  If we define 
\begin{equation}
n=n_{\rm Au}+n_{\rm Ag} \, ;  \, \Delta n =n_{\rm Au}-n_{\rm Ag}
\end{equation}
we can write
\begin{equation}
\Lambda = E_{\rm LAC} - n\bar{\mu}-\Delta n \Delta \mu
\end{equation}
where
\begin{equation}
\bar{\mu} = (\mu_{\rm Au} + \mu_{\rm Ag})/2 \; \; {\rm and} \; \;
\Delta \mu = (\mu_{\rm Au} - \mu_{\rm Ag})/2 \; .
\end{equation}
  The $n \bar{\mu}$ term controls the energetics related to
their having more or less atoms in the chain, regardless of 
species, while the $\Delta n \Delta \mu$ term controls the
energetics related to swapping atoms of different species
in a wire with a given number of atoms. 
  We will confine ourselves here mainly to the latter
behavior considering wires with a constant $n$, and will
consider the free energy
\begin{equation}
\Omega = E_{\rm LAC} - \Delta n \Delta \mu 
= \Lambda + n \bar{\mu} \; .
\end{equation}

  The chemical potentials are defined by the bulk alloy out of which
the wire forms, which constitute the effective reservoirs. 
  In particular, the relative chemical potential can be calculated as
the statistical average energy change when swapping any Au 
atom in the alloy into a Ag atom,
\begin{equation}
\Delta \mu = \left \langle \Delta E ({\rm Au} \rightarrow {\rm Ag})  
\right \rangle \; .
\end{equation}
  This has been obtained by making random substitutions of
Au by Ag in disordered bulk alloy samples of different relative
concentrations, relaxing the structures before and after the 
swap, and computing the energy difference. 
  Face-centered cubic supercell structures with 32 atoms were
used for the purpose.
  The original concentrations of Ag in Au were 0, 7/32, 13/32,
20/32, 26/32, and 1. 
  The results are shown in Fig.~\ref{chemical}.
  Both for $\mu_{\rm Au}$ and $\mu_{\rm Ag}$ the
zero is defined for their respective pure bulk 
crystals.\cite{footrefbulk}
  The curve shows a quite linear behaviour except at
very low Ag concentrations.

\begin{figure}
\includegraphics[scale=0.3]{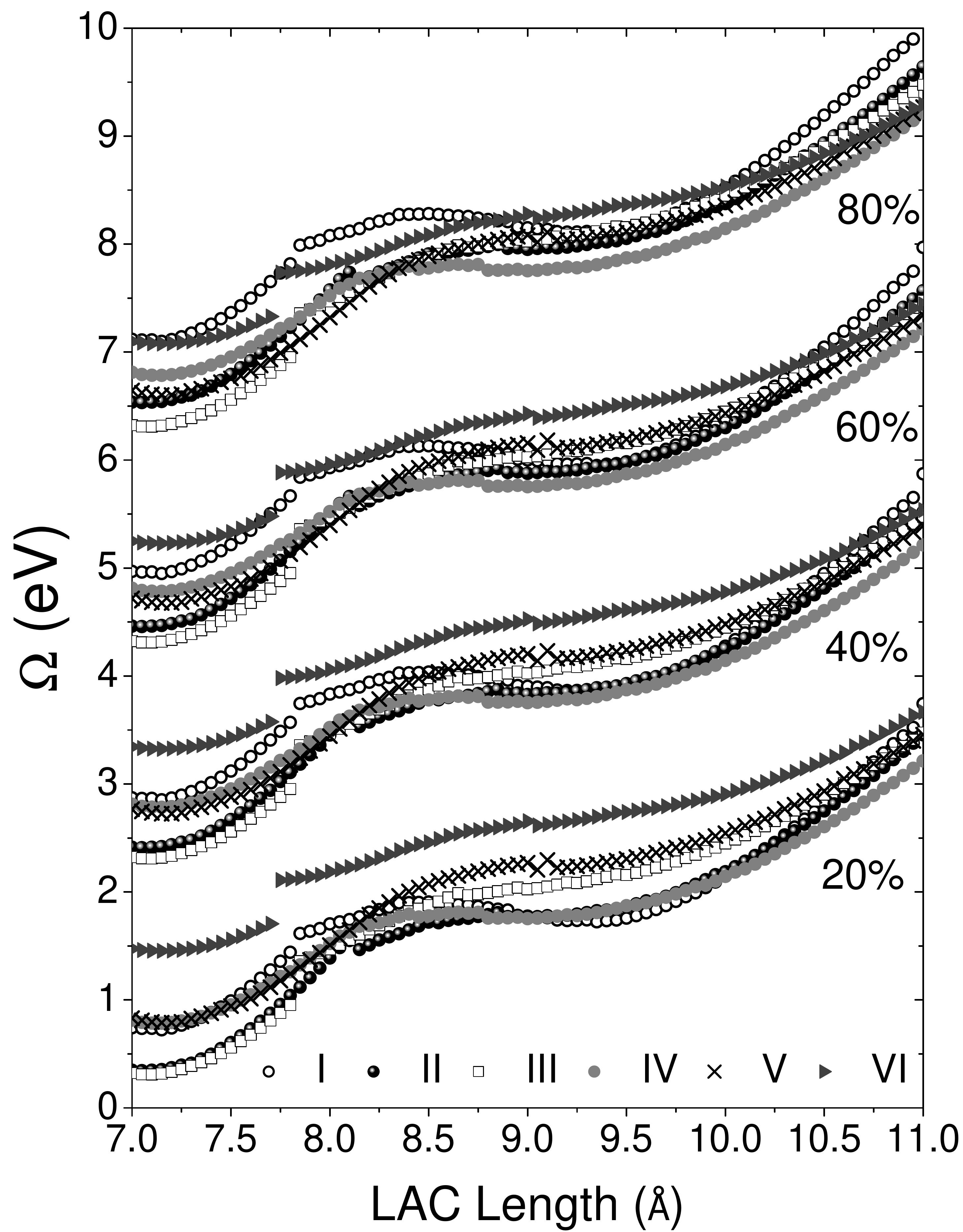}
\caption{\label{OmegaAuAg} 
  Free energy $\Omega$  versus length $L$ for atomic chains 
of different compositions, for different bulk compositions.
  Each group of curves correspond to different bulk alloy 
composition, as specified with the percentages on the
right, and contains six curves, for the wire configurations 
I to VI (see Fig.~\ref{studiedgeometries}), as specified 
in the legend.
  The curves for different bulk compositions have been
shifted upwards by 85.55 eV for convenience.}
\end{figure}

\section{RESULTS}

  Fig.~\ref{OmegaAuAg} shows the free energy $\Omega$ 
as a function of cell length $L$, for the six atomic-wire configurations 
shown in Fig.~\ref{studiedgeometries}, and for the different bulk 
composition of the lead alloys specified on the figure.
  Starting from the left, the regime up to $\sim 7.7$
\AA\ corresponds to zigzag wires in which three consecutive 
atoms define quite an equilateral triangle, i.e., essentially a band 
with two rows of atoms and edge-sharing triangles.
  The regime at larger distances corresponds to the
zigzag structures shown in Fig.~\ref{studiedgeometries}, 
agreeing with what known for Au wires \cite{sanchez1999stiff},
which gradually stretch with growing $L$, until they snap
at $L$-values slightly above what displayed.
  The transitions between regimes, which involve 
``bond" breaking, are discontinuous and hysteretic,
and thus will be quite dependent on quantum and thermal
fluctuations.\cite{hobi2008temperature}  
  The experimentally observed LACs correspond to the intermediate
regime.

  Fig.~\ref{OmegaAuAg} shows that for most alloy 
concentrations the structure IV of Fig.~\ref{studiedgeometries}, 
with Ag-Au alternation is most stable in the LAC regime, with 
higher stabilization for higher Ag content in the bulk alloy.
  This implies that Ag-rich bulk alloys tend to produce
wires with [Ag] tending towards 50\%, i.e. the wires
will be Au enriched.
  These results agree with the mixed and Au enriched
wires seen in the experimental atomically resolved
micrographs.\cite{Bettini2006}
   Reducing the Ag concentration in the alloy there is
a transition to a situation in which the wire could
change equilibrium concentration with stretching,
as seen for 20\%.
  For $L\sim 9.5$ \AA, the wire would tend to pure Au,
thus depleting Ag (although the configuration II with
25\% Ag is close in energy, and indeed lower for 
mechanically unstable shorter $L$).
  Upon stretching, though, the AgAu alternating
wire dominates again, thus replenishing Ag.
 
  In summary, we have defined and calculated from 
first-principles the grand-canonical zero-temperature
free energy governing the fractionation of two species
between an alloyed atomic chain and the bulk alloy.
  We have applied this to Ag/Au alloy chains and 
explained the tendency for Au enrichment in LACs
observed experimentally for Ag rich alloys.
 
\acknowledgments

  PASA acknowledges the funding and support 
of the Erasmus Mundus program and the 
hospitality of the Department of Earth Sciences 
of the University of Cambridge. 
  PASA and DSG acknowledge financial support of 
the Brazilian agencies FAPESP and CNPq.
  They also wish to thank Prof. Daniel Ugarte for many 
helpful discussions.
  EA acknowledges the hospitality of the Institute of 
Physics at the University of Campinas.
  The calculations were done using the CamGrid
high-throughput facility of the University of Cambridge.


\begin{thebibliography}{31}
\expandafter\ifx\csname natexlab\endcsname\relax\def\natexlab#1{#1}\fi
\expandafter\ifx\csname bibnamefont\endcsname\relax
  \def\bibnamefont#1{#1}\fi
\expandafter\ifx\csname bibfnamefont\endcsname\relax
  \def\bibfnamefont#1{#1}\fi
\expandafter\ifx\csname citenamefont\endcsname\relax
  \def\citenamefont#1{#1}\fi
\expandafter\ifx\csname url\endcsname\relax
  \def\url#1{\texttt{#1}}\fi
\expandafter\ifx\csname urlprefix\endcsname\relax\def\urlprefix{URL }\fi
\providecommand{\bibinfo}[2]{#2}
\providecommand{\eprint}[2][]{\url{#2}}

\bibitem[{\citenamefont{Jia et~al.}(2007)\citenamefont{Jia, Shi, Zhao, and
  Wang}}]{Jia2007}
\bibinfo{author}{\bibfnamefont{J.}~\bibnamefont{Jia}},
  \bibinfo{author}{\bibfnamefont{D.}~\bibnamefont{Shi}},
  \bibinfo{author}{\bibfnamefont{J.}~\bibnamefont{Zhao}}, \bibnamefont{and}
  \bibinfo{author}{\bibfnamefont{B.}~\bibnamefont{Wang}},
  \bibinfo{journal}{Phys. Rev. B} \textbf{\bibinfo{volume}{76}},
  \bibinfo{pages}{165420} (\bibinfo{year}{2007}).

\bibitem[{\citenamefont{Agrait et~al.}(2003)\citenamefont{Agrait, Yeyati, and
  van Ruitenbeek}}]{Agrait2003}
\bibinfo{author}{\bibfnamefont{N.}~\bibnamefont{Agrait}},
  \bibinfo{author}{\bibfnamefont{A.~L.} \bibnamefont{Yeyati}},
  \bibnamefont{and} \bibinfo{author}{\bibfnamefont{J.~M.} \bibnamefont{van
  Ruitenbeek}}, \bibinfo{journal}{Phys. Rep.} p.~\bibinfo{pages}{81}
  (\bibinfo{year}{2003}).

\bibitem[{\citenamefont{Yanson and Yanson}(1999)}]{Yanson1999}
\bibinfo{author}{\bibfnamefont{A.}~\bibnamefont{Yanson}} \bibnamefont{and}
  \bibinfo{author}{\bibfnamefont{I.}~\bibnamefont{Yanson}},
  \bibinfo{journal}{Nature} \textbf{\bibinfo{volume}{400}},
  \bibinfo{pages}{144} (\bibinfo{year}{1999}).

\bibitem[{\citenamefont{Mares and Van~Ruitenbeek}(2005)}]{Mares2005}
\bibinfo{author}{\bibfnamefont{A.}~\bibnamefont{Mares}} \bibnamefont{and}
  \bibinfo{author}{\bibfnamefont{J.}~\bibnamefont{Van~Ruitenbeek}},
  \bibinfo{journal}{Phys. Rev. B} \textbf{\bibinfo{volume}{72}},
  \bibinfo{pages}{205402} (\bibinfo{year}{2005}).

\bibitem[{\citenamefont{Rodrigues et~al.}(2000)\citenamefont{Rodrigues, Fuhrer,
  and Ugarte}}]{Rodrigues2000}
\bibinfo{author}{\bibfnamefont{V.}~\bibnamefont{Rodrigues}},
  \bibinfo{author}{\bibfnamefont{T.}~\bibnamefont{Fuhrer}}, \bibnamefont{and}
  \bibinfo{author}{\bibfnamefont{D.}~\bibnamefont{Ugarte}},
  \bibinfo{journal}{Phys. Rev. Lett.} \textbf{\bibinfo{volume}{85}},
  \bibinfo{pages}{4124} (\bibinfo{year}{2000}).

\bibitem[{\citenamefont{Kondo and Takayanagi}(1997)}]{Kondo1997}
\bibinfo{author}{\bibfnamefont{Y.}~\bibnamefont{Kondo}} \bibnamefont{and}
  \bibinfo{author}{\bibfnamefont{K.}~\bibnamefont{Takayanagi}},
  \bibinfo{journal}{Phys. Rev. Lett.} \textbf{\bibinfo{volume}{79}},
  \bibinfo{pages}{3455} (\bibinfo{year}{1997}).

\bibitem[{\citenamefont{Kizuka}(1998)}]{Kizuka1998}
\bibinfo{author}{\bibfnamefont{T.}~\bibnamefont{Kizuka}},
  \bibinfo{journal}{Phys. Rev. Lett.} \textbf{\bibinfo{volume}{81}},
  \bibinfo{pages}{4448} (\bibinfo{year}{1998}).

\bibitem[{\citenamefont{Kizuka et~al.}(1997)\citenamefont{Kizuka, Yamada,
  Deguchi, Naruse, and Tanaka}}]{Kizuka1997}
\bibinfo{author}{\bibfnamefont{T.}~\bibnamefont{Kizuka}},
  \bibinfo{author}{\bibfnamefont{K.}~\bibnamefont{Yamada}},
  \bibinfo{author}{\bibfnamefont{S.}~\bibnamefont{Deguchi}},
  \bibinfo{author}{\bibfnamefont{M.}~\bibnamefont{Naruse}}, \bibnamefont{and}
  \bibinfo{author}{\bibfnamefont{N.}~\bibnamefont{Tanaka}},
  \bibinfo{journal}{Phys. Rev. B} \textbf{\bibinfo{volume}{55}},
  \bibinfo{pages}{7398} (\bibinfo{year}{1997}).

\bibitem[{\citenamefont{Koizumi et~al.}(2001)\citenamefont{Koizumi, Oshima,
  Kondo, and Takayanagi}}]{Koizumi2001}
\bibinfo{author}{\bibfnamefont{H.}~\bibnamefont{Koizumi}},
  \bibinfo{author}{\bibfnamefont{Y.}~\bibnamefont{Oshima}},
  \bibinfo{author}{\bibfnamefont{Y.}~\bibnamefont{Kondo}}, \bibnamefont{and}
  \bibinfo{author}{\bibfnamefont{K.}~\bibnamefont{Takayanagi}},
  \bibinfo{journal}{Ultramicroscopy} \textbf{\bibinfo{volume}{88}},
  \bibinfo{pages}{17} (\bibinfo{year}{2001}).

\bibitem[{\citenamefont{Ohnishi et~al.}(1998)\citenamefont{Ohnishi, Kondo, and
  Takayanagi}}]{Ohnishi1998}
\bibinfo{author}{\bibfnamefont{H.}~\bibnamefont{Ohnishi}},
  \bibinfo{author}{\bibfnamefont{Y.}~\bibnamefont{Kondo}}, \bibnamefont{and}
  \bibinfo{author}{\bibfnamefont{K.}~\bibnamefont{Takayanagi}},
  \bibinfo{journal}{Nature} \textbf{\bibinfo{volume}{395}},
  \bibinfo{pages}{780} (\bibinfo{year}{1998}).

\bibitem[{\citenamefont{Rodrigues and Ugarte}(2001)}]{Rodrigues2001}
\bibinfo{author}{\bibfnamefont{V.}~\bibnamefont{Rodrigues}} \bibnamefont{and}
  \bibinfo{author}{\bibfnamefont{D.}~\bibnamefont{Ugarte}},
  \bibinfo{journal}{Phys. Rev. B} \textbf{\bibinfo{volume}{6307}}
  (\bibinfo{year}{2001}).

\bibitem[{\citenamefont{Lagos et~al.}(2012)\citenamefont{Lagos, Autreto,
  Galvao, and Ugarte}}]{lagos2012correlation}
\bibinfo{author}{\bibfnamefont{M.}~\bibnamefont{Lagos}},
  \bibinfo{author}{\bibfnamefont{P.}~\bibnamefont{Autreto}},
  \bibinfo{author}{\bibfnamefont{D.}~\bibnamefont{Galvao}}, \bibnamefont{and}
  \bibinfo{author}{\bibfnamefont{D.}~\bibnamefont{Ugarte}},
  \bibinfo{journal}{J. Appl. Phys.} \textbf{\bibinfo{volume}{111}},
  \bibinfo{pages}{124316} (\bibinfo{year}{2012}).

\bibitem[{\citenamefont{Lagos et~al.}(2009)\citenamefont{Lagos, Sato, Bettini,
  Rodrigues, Galvao, and Ugarte}}]{Lagos2009}
\bibinfo{author}{\bibfnamefont{M.}~\bibnamefont{Lagos}},
  \bibinfo{author}{\bibfnamefont{F.}~\bibnamefont{Sato}},
  \bibinfo{author}{\bibfnamefont{J.}~\bibnamefont{Bettini}},
  \bibinfo{author}{\bibfnamefont{V.}~\bibnamefont{Rodrigues}},
  \bibinfo{author}{\bibfnamefont{D.}~\bibnamefont{Galvao}}, \bibnamefont{and}
  \bibinfo{author}{\bibfnamefont{D.}~\bibnamefont{Ugarte}},
  \bibinfo{journal}{Nature Nanotechnology} \textbf{\bibinfo{volume}{4}},
  \bibinfo{pages}{149} (\bibinfo{year}{2009}).

\bibitem[{\citenamefont{Kondo and Takayanagi}(2000)}]{Kondo2000}
\bibinfo{author}{\bibfnamefont{Y.}~\bibnamefont{Kondo}} \bibnamefont{and}
  \bibinfo{author}{\bibfnamefont{K.}~\bibnamefont{Takayanagi}},
  \bibinfo{journal}{Science} \textbf{\bibinfo{volume}{289}},
  \bibinfo{pages}{606} (\bibinfo{year}{2000}).

\bibitem[{\citenamefont{Oshima et~al.}(2002)\citenamefont{Oshima, Koizumi,
  Mouri, Hirayama, Takayanagi, and Kondo}}]{Oshima2002}
\bibinfo{author}{\bibfnamefont{Y.}~\bibnamefont{Oshima}},
  \bibinfo{author}{\bibfnamefont{H.}~\bibnamefont{Koizumi}},
  \bibinfo{author}{\bibfnamefont{K.}~\bibnamefont{Mouri}},
  \bibinfo{author}{\bibfnamefont{H.}~\bibnamefont{Hirayama}},
  \bibinfo{author}{\bibfnamefont{K.}~\bibnamefont{Takayanagi}},
  \bibnamefont{and} \bibinfo{author}{\bibfnamefont{Y.}~\bibnamefont{Kondo}},
  \bibinfo{journal}{Phys. Rev. B} \textbf{\bibinfo{volume}{65}},
  \bibinfo{pages}{121401} (\bibinfo{year}{2002}).

\bibitem[{\citenamefont{Oshima et~al.}(2003)\citenamefont{Oshima, Onga, and
  Takayanagi}}]{Oshima2003}
\bibinfo{author}{\bibfnamefont{Y.}~\bibnamefont{Oshima}},
  \bibinfo{author}{\bibfnamefont{A.}~\bibnamefont{Onga}}, \bibnamefont{and}
  \bibinfo{author}{\bibfnamefont{K.}~\bibnamefont{Takayanagi}},
  \bibinfo{journal}{Phys. Rev. Lett.} \textbf{\bibinfo{volume}{91}},
  \bibinfo{pages}{205503} (\bibinfo{year}{2003}).

\bibitem[{\citenamefont{Coura et~al.}(2004)\citenamefont{Coura, Legoas,
  Moreira, Sato, Rodrigues, Dantas, Ugarte, and
  Galv{\~a}o}}]{coura2004structural}
\bibinfo{author}{\bibfnamefont{P.}~\bibnamefont{Coura}},
  \bibinfo{author}{\bibfnamefont{S.}~\bibnamefont{Legoas}},
  \bibinfo{author}{\bibfnamefont{A.}~\bibnamefont{Moreira}},
  \bibinfo{author}{\bibfnamefont{F.}~\bibnamefont{Sato}},
  \bibinfo{author}{\bibfnamefont{V.}~\bibnamefont{Rodrigues}},
  \bibinfo{author}{\bibfnamefont{S.}~\bibnamefont{Dantas}},
  \bibinfo{author}{\bibfnamefont{D.}~\bibnamefont{Ugarte}}, \bibnamefont{and}
  \bibinfo{author}{\bibfnamefont{D.}~\bibnamefont{Galv{\~a}o}},
  \bibinfo{journal}{Nano Letters} \textbf{\bibinfo{volume}{4}},
  \bibinfo{pages}{1187} (\bibinfo{year}{2004}).

\bibitem[{\citenamefont{Nakamura et~al.}(1999)\citenamefont{Nakamura,
  Brandbyge, Hansen, and Jacobsen}}]{nakamura1999density}
\bibinfo{author}{\bibfnamefont{A.}~\bibnamefont{Nakamura}},
  \bibinfo{author}{\bibfnamefont{M.}~\bibnamefont{Brandbyge}},
  \bibinfo{author}{\bibfnamefont{L.}~\bibnamefont{Hansen}}, \bibnamefont{and}
  \bibinfo{author}{\bibfnamefont{K.}~\bibnamefont{Jacobsen}},
  \bibinfo{journal}{Phys. Rev. Lett.} \textbf{\bibinfo{volume}{82}},
  \bibinfo{pages}{1538} (\bibinfo{year}{1999}).

\bibitem[{\citenamefont{S{\'a}nchez-Portal
  et~al.}(1999)\citenamefont{S{\'a}nchez-Portal, Artacho, Junquera,
  Ordej{\'o}n, Garc{\'\i}a, and Soler}}]{sanchez1999stiff}
\bibinfo{author}{\bibfnamefont{D.}~\bibnamefont{S{\'a}nchez-Portal}},
  \bibinfo{author}{\bibfnamefont{E.}~\bibnamefont{Artacho}},
  \bibinfo{author}{\bibfnamefont{J.}~\bibnamefont{Junquera}},
  \bibinfo{author}{\bibfnamefont{P.}~\bibnamefont{Ordej{\'o}n}},
  \bibinfo{author}{\bibfnamefont{A.}~\bibnamefont{Garc{\'\i}a}},
  \bibnamefont{and} \bibinfo{author}{\bibfnamefont{J.}~\bibnamefont{Soler}},
  \bibinfo{journal}{Phys. Rev. Lett.} \textbf{\bibinfo{volume}{83}},
  \bibinfo{pages}{3884} (\bibinfo{year}{1999}).

\bibitem[{\citenamefont{Senger et~al.}(2004)\citenamefont{Senger, Dag, and
  Ciraci}}]{senger2004chiral}
\bibinfo{author}{\bibfnamefont{R.}~\bibnamefont{Senger}},
  \bibinfo{author}{\bibfnamefont{S.}~\bibnamefont{Dag}}, \bibnamefont{and}
  \bibinfo{author}{\bibfnamefont{S.}~\bibnamefont{Ciraci}},
  \bibinfo{journal}{Phys. Rev. Lett.} \textbf{\bibinfo{volume}{93}},
  \bibinfo{pages}{196807} (\bibinfo{year}{2004}).

\bibitem[{\citenamefont{Tosatti et~al.}(2001)\citenamefont{Tosatti, Prestipino,
  Kostlmeier, Dal~Corso, and Di~Tolla}}]{tosatti2001string}
\bibinfo{author}{\bibfnamefont{E.}~\bibnamefont{Tosatti}},
  \bibinfo{author}{\bibfnamefont{S.}~\bibnamefont{Prestipino}},
  \bibinfo{author}{\bibfnamefont{S.}~\bibnamefont{Kostlmeier}},
  \bibinfo{author}{\bibfnamefont{A.}~\bibnamefont{Dal~Corso}},
  \bibnamefont{and} \bibinfo{author}{\bibfnamefont{F.}~\bibnamefont{Di~Tolla}},
  \bibinfo{journal}{Science} \textbf{\bibinfo{volume}{291}},
  \bibinfo{pages}{288} (\bibinfo{year}{2001}).

\bibitem[{\citenamefont{Bettini et~al.}(2006)\citenamefont{Bettini, Sato,
  Coura, Dantas, Galv{\~a}o, and Ugarte}}]{Bettini2006}
\bibinfo{author}{\bibfnamefont{J.}~\bibnamefont{Bettini}},
  \bibinfo{author}{\bibfnamefont{F.}~\bibnamefont{Sato}},
  \bibinfo{author}{\bibfnamefont{P.}~\bibnamefont{Coura}},
  \bibinfo{author}{\bibfnamefont{S.}~\bibnamefont{Dantas}},
  \bibinfo{author}{\bibfnamefont{D.}~\bibnamefont{Galv{\~a}o}},
  \bibnamefont{and} \bibinfo{author}{\bibfnamefont{D.}~\bibnamefont{Ugarte}},
  \bibinfo{journal}{Nature Nanotechnology} \textbf{\bibinfo{volume}{1}},
  \bibinfo{pages}{182} (\bibinfo{year}{2006}).

\bibitem[{\citenamefont{Fa and Dong}(2008)}]{fa2008stability}
\bibinfo{author}{\bibfnamefont{W.}~\bibnamefont{Fa}} \bibnamefont{and}
  \bibinfo{author}{\bibfnamefont{J.}~\bibnamefont{Dong}}, \bibinfo{journal}{
  J. Chem. Phys.} \textbf{\bibinfo{volume}{128}},
  \bibinfo{pages}{244703} (\bibinfo{year}{2008}).

\bibitem[{\citenamefont{Ordej{\'o}n et~al.}(1996)\citenamefont{Ordej{\'o}n,
  Artacho, and Soler}}]{Ordejon1996}
\bibinfo{author}{\bibfnamefont{P.}~\bibnamefont{Ordej{\'o}n}},
  \bibinfo{author}{\bibfnamefont{E.}~\bibnamefont{Artacho}}, \bibnamefont{and}
  \bibinfo{author}{\bibfnamefont{J.}~\bibnamefont{Soler}},
  \bibinfo{journal}{Phys. Rev. B} \textbf{\bibinfo{volume}{53}},
  \bibinfo{pages}{10441} (\bibinfo{year}{1996}).

\bibitem[{\citenamefont{Soler et~al.}(2002)\citenamefont{Soler, Artacho, Gale,
  Garc{\'\i}a, Junquera, Ordej{\'o}n, and S{\'a}nchez-Portal}}]{Soler2002}
\bibinfo{author}{\bibfnamefont{J.}~\bibnamefont{Soler}},
  \bibinfo{author}{\bibfnamefont{E.}~\bibnamefont{Artacho}},
  \bibinfo{author}{\bibfnamefont{J.}~\bibnamefont{Gale}},
  \bibinfo{author}{\bibfnamefont{A.}~\bibnamefont{Garc{\'\i}a}},
  \bibinfo{author}{\bibfnamefont{J.}~\bibnamefont{Junquera}},
  \bibinfo{author}{\bibfnamefont{P.}~\bibnamefont{Ordej{\'o}n}},
  \bibnamefont{and}
  \bibinfo{author}{\bibfnamefont{D.}~\bibnamefont{S{\'a}nchez-Portal}},
  \bibinfo{journal}{Journal of Physics: Condensed Matter}
  \textbf{\bibinfo{volume}{14}}, \bibinfo{pages}{2745} (\bibinfo{year}{2002}).

\bibitem[{\citenamefont{Perdew et~al.}(1996)\citenamefont{Perdew, Burke, and
  Ernzerhof}}]{Perdew1996}
\bibinfo{author}{\bibfnamefont{J.}~\bibnamefont{Perdew}},
  \bibinfo{author}{\bibfnamefont{K.}~\bibnamefont{Burke}}, \bibnamefont{and}
  \bibinfo{author}{\bibfnamefont{M.}~\bibnamefont{Ernzerhof}},
  \bibinfo{journal}{Phys. Rev. Lett.} \textbf{\bibinfo{volume}{77}},
  \bibinfo{pages}{3865} (\bibinfo{year}{1996}).

\bibitem[{\citenamefont{Troullier and Martins}(1991)}]{Troullier1991}
\bibinfo{author}{\bibfnamefont{N.}~\bibnamefont{Troullier}} \bibnamefont{and}
  \bibinfo{author}{\bibfnamefont{J.}~\bibnamefont{Martins}},
  \bibinfo{journal}{Phys. Rev. B} \textbf{\bibinfo{volume}{43}},
  \bibinfo{pages}{1993} (\bibinfo{year}{1991}).

\bibitem[{\citenamefont{Anglada et~al.}(2002)\citenamefont{Anglada, M.~Soler,
  Junquera, and Artacho}}]{Anglada2002}
\bibinfo{author}{\bibfnamefont{E.}~\bibnamefont{Anglada}},
  \bibinfo{author}{\bibfnamefont{J.}~\bibnamefont{M.~Soler}},
  \bibinfo{author}{\bibfnamefont{J.}~\bibnamefont{Junquera}}, \bibnamefont{and}
  \bibinfo{author}{\bibfnamefont{E.}~\bibnamefont{Artacho}},
  \bibinfo{journal}{Phys. Rev. B} \textbf{\bibinfo{volume}{66}},
  \bibinfo{pages}{205101} (\bibinfo{year}{2002}).

\bibitem[{\citenamefont{Monkhorst and Pack}(1976)}]{Monkhorst1976}
\bibinfo{author}{\bibfnamefont{H.}~\bibnamefont{Monkhorst}} \bibnamefont{and}
  \bibinfo{author}{\bibfnamefont{J.}~\bibnamefont{Pack}},
  \bibinfo{journal}{Phys. Rev. B} \textbf{\bibinfo{volume}{13}},
  \bibinfo{pages}{5188} (\bibinfo{year}{1976}).

\bibitem[{\citenamefont{Hobi~Jr et~al.}(2008)\citenamefont{Hobi~Jr, Fazzio, and
  da~Silva}}]{hobi2008temperature}
\bibinfo{author}{\bibfnamefont{E.}~\bibnamefont{Hobi~Jr}},
  \bibinfo{author}{\bibfnamefont{A.}~\bibnamefont{Fazzio}}, \bibnamefont{and}
  \bibinfo{author}{\bibfnamefont{A.}~\bibnamefont{da~Silva}},
  \bibinfo{journal}{Phys. Rev. Lett.} \textbf{\bibinfo{volume}{100}},
  \bibinfo{pages}{56104} (\bibinfo{year}{2008}).

\bibitem[{\citenamefont{Frenkel and Smit}(2002)}]{frenkel2002understanding}
\bibinfo{author}{\bibfnamefont{D.}~\bibnamefont{Frenkel}} \bibnamefont{and}
  \bibinfo{author}{\bibfnamefont{B.}~\bibnamefont{Smit}},
  \emph{\bibinfo{title}{Understanding molecular simulation: from algorithms to
  applications}}, vol.~\bibinfo{volume}{1} (\bibinfo{publisher}{Academic Pr},
  \bibinfo{year}{2002}).

 \bibitem{footrefbulk}
This is for convenience with the numbers. 
The bare $\Delta \mu$ coming from 
$\Delta E^{\rm DFT}({\rm Au} \rightarrow {\rm Ag})$
gives large numbers referring to the separate cores and
electrons. Here we use 
$\Delta E^{\rm DFT}({\rm Au} \rightarrow {\rm Ag}) 
+ \mu_{\rm Au}^b - \mu_{\rm Ag}^b$, the two latter terms being
the DFT energy per atom of bulk Au and bulk Ag, respectively.
The same is done in the definition of $E_{\rm LAC}$.

\end{thebibliography}
\end{document}